\documentclass[]{emulateapj}
\usepackage{psfig}

\def\etal{{et\,al. }}
\def\msun{M$_{\odot}$}

\def\degs{\ifmmode ^{\circ}\else$^{\circ}$\fi}
\def\amin{\ifmmode ^{\prime}\else$^{\prime}$\fi}
\def\asec{\ifmmode ^{\prime\prime}\else$^{\prime\prime}$\fi}
\newbox\grsign \setbox\grsign=\hbox{$>$}
\newdimen\grdimen \grdimen=\ht\grsign
\newbox\laxbox \newbox\gaxbox
\setbox\gaxbox=\hbox{\raise.5ex\hbox{$>$}\llap
     {\lower.5ex\hbox{$\sim$}}}\ht1=\grdimen\dp1=0pt
\setbox\laxbox=\hbox{\raise.5ex\hbox{$<$}\llap
     {\lower.5ex\hbox{$\sim$}}}\ht2=\grdimen\dp2=0pt

\shorttitle{Scaling relations for nuclei of Seyfert 1 galaxies}
\shortauthors{Salvato, Greiner \& Kuhlbrodt}

\begin{document}


\title{ Multiwavelength scaling relations for nuclei of Seyfert galaxies\thanks{Partly based on observations collected at the European Southern Observatory, La Silla, Chile and at Calar Alto, Spain.}}


\author{M. Salvato\altaffilmark{1,2} and J. Greiner\altaffilmark{1,2}}
\affil{MPE, Giessenbachstr. 1, 85748 Garching, Germany}
\altaffiltext{2}{Formerly at AIP, An der Sternwarte 16, 14482 Potsdam, Germany}

\author{B. Kuhlbrodt\altaffilmark{3}}
\affil{Hamburg Observatory, Gojenbergsweg 112, 21029 Hamburg, Germany}






\begin{abstract}
We analyze an X-ray flux-limited, complete sample of 93 AGN at z$<
$0.1, selected from the ROSAT Bright Survey. Two thirds of the sample
are Seyfert 1 galaxies (Sy1) and one third are Narrow-Line Seyfert 1
galaxies (NLSy1). We have obtained optical images of all objects. By
modeling the host galaxy and the AGN central component we decompose
the optical emission into nuclear, bulge and disk components,
respectively.  We find that the nuclear optical luminosity, thought to
be associated with the accretion disk surrounding the active black
hole, correlates with the X-ray luminosity, the radio luminosity and
the black hole mass.
\end{abstract}




\section{Introduction}

The last few years have brought fundamental progress in understanding
the suspected connection between active galactic nuclei (AGN),
characterized by non-thermal emission from a massive ($10^6 - 10^9$
\msun) black hole (BH), and normal galaxies, whose extended emission
is thermal.  In particular, now 1) there is robust evidence that
quasars reside in galaxies (Bahcall et al. 1997); 2) it is established
that massive ``quiescent'' BHs are also present in the central regions
of normal galaxies (Magorrian et al. 1998); 3) a statistical
correlation between BH mass and mass or velocity dispersion of the
host galaxy's bulge exists (Gebhardt et al. 2000; Ferrarese \& Merritt
2000), which suggests a common evolution for galaxies and their
central, massive BHs; 4) the previous correlations hold for normal as
well as AGN-host galaxies (McLure \& Dunlop 2002).

For AGN with typical masses of $10^6 - 10^9$ \msun, the bulk of the
accretion radiation is emitted in the soft X-ray and EUV/UV range.  A
fraction of this optically thick radiation is Compton-upscattered by a
corona of electrons surrounding the BH. This process is thought to
produce the typical power law X-ray spectra observed in all AGN. The
primary soft X-ray/EUV radiation of the accretion disk also ionizes
the surrounding gas, producing the well-known AGN emission lines. At
radio frequencies, the core luminosity of the AGN is due to
synchrotron emission by relativistic electrons immersed in a
seemingly global magnetic field.  It is conceivable that these
electrons are the same which Compton-upscatter the disk photons. It is
therefore interesting to ask whether there is a direct correlation
between radio, optical and X-ray emission in the central part of the
AGN where the accretion disk is located.

While information in the radio and X-ray bands is already available
from all-sky surveys like the NRAO VLA Sky Survey (NVSS, Condon \etal
1998), FIRST (Faint Images of the Radio Sky at Twenty centimeters;
Becker \etal 1995) and the ROSAT Bright Survey (RBS, Schwope \etal
2000), optical data for the central kiloparsec of AGN are more
difficult to obtain.  Indeed, in modeling AGN and their host galaxies,
the information contained in the central pixels has been usually
rejected (see e.g. Virani \etal 2000).  With a new well-tested,
well-performing and robust fitting algorithm (Kuhlbrodt \etal 2001),
it has recently been possible to model both the host galaxy {\em and}
the nucleus from optical images.

Here we apply this decomposition algorithm to a new sample of objects,
and use the results of the decomposition to study the relations
between the properties of the central point source, in particular the
X-ray ($0.5-2.0$ keV), optical (R band) and radio (1.49 GHz) emission.
After presenting the sample selection in \S 1 and the catalog
data in \S 2, the relations between the different
parameters are described in \S 3, and discussed in \S 4. Unless
specified differently, a value of H$_o$=50 km Mpc$^{-1}$s$^{-1}$ has
been used throughout this paper.

\section{Construction of the sample}

\begin{figure}[th]
\begin{center}
 \vbox{\psfig{figure=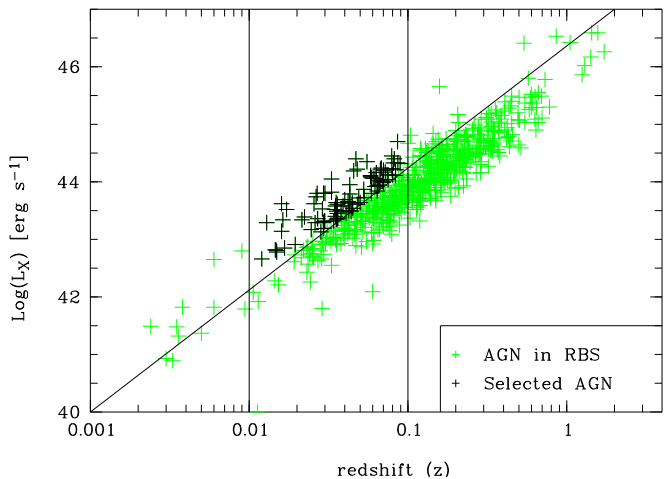,width=8.8cm,angle=270,clip=}}\par
  \caption[]{AGN in the RBS catalogue. Solid lines define the 
selection criteria in redshift (0.009$<$z$<$0.1; vertical lines) 
and in flux (cr $>0.3$ cts/s; diagonal). Black symbols form the 
sample used in this work.}
\label{Lxz}
\end{center}
\end{figure}

We used the ROSAT Bright Survey (RBS, Schwope \etal 2000), consisting
of the $\sim$ 2000 brightest, optically identified sources detected
during the ROSAT All Sky Survey (RASS, Voges \etal 1999).  Selecting
at X-rays has the advantage that the contribution of the emission from
the host galaxy is small with respect to the AGN.  We selected the
strongest (ROSAT PSPC count rate (cr) above 0.3 cts s$^{-1}$ in the
0.5--2.0 keV range), local (0.009$<$z$<$0.1) AGN in order to achieve
reasonable spatial resolution with ground-based optical observations.
The sample consists of 93 AGN (black symbols in Fig. \ref{Lxz}).  At
the time when the work started, all of them were classified in
NED\footnote{ NASA/IPAC Extragalactic Database} as Sy1 galaxies, but
in the meantime about 30\% of them have been found to be NLSy1
(Veron-Cetty \& Veron 2001). The size of the sample thus allows us to
work with 2 independent, complete samples, Sy1 (64) and NLSy1 (29),
with comparable X-ray luminosity.  In both cases the observer has a
direct view of the AGN, i.e. unobstructed by the torus.
\section{Data collection}
All optical data was acquired in dedicated observing runs, described
in detail elsewhere (Salvato 2002; Salvato 2003, in preparation). Here
we give only some brief characteristics.  In order to explore the
relationship of nuclear luminosities in various wavelengths we
collected data from public catalogs for the X-ray and radio
wavelengths.\\

{\it Optical data.}
The entire sample has been observed in the R band with a typical
exposure time of 10 minutes at the 1.5m Danish telescope (Bessel R
filter) at La Silla and of 20 minutes at the 1.2m telescope (Johnson R
filter) at Calar Alto. We have applied the two-dimensional
decomposition algorithm (Kuhlbrodt \etal 2001) to all images, which
takes into account the point spread function (PSF) of the
corresponding image. A three-component model has been applied,
involving a point-like nuclear component, a bulge component with a
R$^{1/4}$ radial dependence, and an exponential disk
component. Besides the spatial parameters we therefore obtain
magnitudes of these three components.  These derived magnitudes have
been corrected for Galactic absorption, using the maps of Dickey \&
Lockman (1990) and converted to the visual extinction as derived by
Predehl \& Schmitt (1995). A comparison of our fit parameters (colors,
disk and bulge parameters, etc.) with a sub-sample of galaxies studied
already earlier (e.g. McLure \& Dunlop 2001), reveals perfect
agreement (details in Salvato 2002), and thus provides some confidence
that also the nuclear light component is fitted correctly. {\it In \S
4, this optical nuclear luminosity is supposed to represent the
accretion disk surrounding the BH.}\\


{\it Black hole masses.}
BH masses are computed from the derived bulge luminosity,
using a unique equation for Sy1 and NLSy1 as in McLure \& Dunlop
(2002): $log(M_{\rm BH}$/{\rm \msun })= $-0.50(\pm0.02) M_{\rm R} -
2.96(\pm 0.48)$ which is valid under the assumption that differences
between Sy1 and NLSy1 are due to geometrical rather than physical
effects.  Our masses (Tab. \ref{Kaspi}), after correcting for the
different cosmology, agree well with those derived by reverberation
mapping (Kaspi \etal 2000) for the common sources.\\


{\it X-ray luminosity.}
Typically, non-active galaxies have very low X-ray luminosities; thus
the X-ray emission at a ROSAT/PSPC count rate above 0.3 cts/s is
likely produced only in the nuclei of AGN. However, the X-ray emission
is not directly related to the accretion disk, because due to the
canonical temperatures of the disk emission peaks in the UV.  The
X-ray luminosity has been computed using the X-ray flux in the
0.5--2.0 keV (Tab. 2 in Schwope \etal 2000), assuming a power law
spectrum with photon index of --2 and applying the corresponding
Galactic absorption and $k$ correction. We caution that this approach
yields unreliable and inconsistent results for few NLSy1 in our
sample, having very steep, soft X-ray spectra.\\


{\it Radio data.}
We cross-correlated our sample with the NVSS and FIRST radio
catalogues, and, owing to their limited sky coverage, 53 matches were
found.  Both surveys are centered at 1.49 GHz ($\lambda$ = 20 cm),
where the synchrotron emission of our targets may be associated with
the nuclear activity but also with the spatially-extended star
formation activity, if any.  NVSS and FIRST map the sky at different
spatial resolution (45\asec and 5\asec, respectively), and thus allow
to disentangle the two potential contributions.  In fact, if the
unresolved AGN produces all the radio emission, the fluxes listed in
the two catalogues must be consistent; otherwise the NVSS flux must be
larger.  A flux-flux plot shows that galaxies with and without the
extra contribution to the radio emission from star formation activity
can be separated by a line of equation $\log(L_{1.49 {\rm GHz}}({\rm
NVSS})= \log(L_{1.49 {\rm GHz}}({\rm FIRST}))+0.3$.  This selection
criterion is robust against a maximal 20\% variability of the AGN
radio flux (Falcke \etal 2001).  Thus the radio luminosities of 21
out of 53 galaxies are attributed only to their nuclear activity.


 \begin{table}[th]
\begin{center}
\caption{Comparison of BH masses obtained following McLure\&Dunlop (2002) and
 via reverberation mapping (Kaspi \etal 2000). All values are computed
  using H$_0$=75 km Mpc$^{-1}$s$^{-1}$.}
\begin{tabular}{lcc}
\hline{\smallskip}
Source Name & Kaspi \etal  & This work\\
     &  $\log(M_{bh}/$\msun) &$\log(M_{bh}$/\msun)\\
\noalign{\smallskip}\hline\hline
n011 (Fairall 9) & 7.59$\div$ 8.01 &7.89$\div$8.83 \\
n054 (PG 1211+14)& 7.45  $\div$ 7.7 & 8.44$\div$ 9.34\\  
n065 (IC 4329A)  & 7.25 $\downarrow$&  6.76  $\div$7.78\\   
n069 (NGC 5548)  & 8.02 $\div $8.16 & 7.14$\div$ 8.12\\  
\noalign{\smallskip}\hline
\end{tabular}
\label{Kaspi}
\end{center}
\vspace*{-0.3cm}
\end{table}

\begin{table}[ht]
\begin{center}
\caption{Coefficients for the equations 1 and 2.  The Spearman Rank
correlation coefficients $\rho$ , are given in col. 3 and 6.}
\begin{tabular}{lcccccc}
\hline{\smallskip}
 &\multicolumn{3}{c}{Equation 1} &\multicolumn{3}{c}{Equation 2}\\
     &  a &b&$\rho$& c&d&$\rho$\\
\noalign{\smallskip}\hline\hline
$\!\!$All   & 0.58$\pm$0.06 & 18.44$\pm$2.6 &0.74&0.96$\pm$0.24 & -4.18$\pm$10.7&0.68\\
$\!\!$Sy1   & 0.56$\pm$0.07 & 19.48$\pm$3.0 &0.72&0.74$\pm$0.22 & 4.80$\pm$9.6&0.60\\
$\!\!$NLSy1 & 0.61$\pm$0.11 & 17.09$\pm$4.9 &0.82&1.18$\pm$0.52 &$\!\!$ -13.95$\pm$22.9&0.92$\!\!$\\
\noalign{\smallskip}\hline
\end{tabular}
\label{ab}
\end{center}
\end{table}


\section{Correlation analysis}

\begin{figure}[th]
\begin{center}
\hspace{-0.5cm}
 \vbox{\psfig{figure=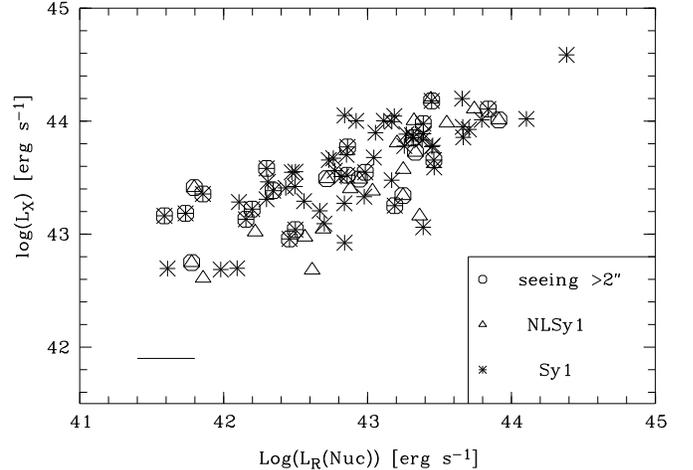,width=8.8cm,angle=-90,clip=}}\par
\caption[]{X-ray versus optical (R-band) luminosity for the sample.
  Asterisks (open triangles) represent Sy1 (NLSy1) galaxies.  For 30
  out of the 93 objects imaging was obtained under bad seeing
  condition ($>$2\asec). These are marked by open circles around their
  respective symbols and were ignored for determining the linear
  correlation between the X-ray and optical luminosities.  Larger
  seeing implies more flux in the nuclear component relative to the
  bulge component, and therefore those encircled data points have to
  be treated as upper limits on the nuclear optical emission.  At
  bottom left the expected error for the nuclear, optical luminosity
  is given. Sy1 and NLSy1 behave in the same way.}
\label{lpsflx}
\end{center}
\end{figure}

{\it Optical vs. X-ray emission.}
A strong linear relation is found between the optical accretion disk
and the X-ray luminosity (Fig. \ref{lpsflx}), of the form
\begin{equation}
\log(L_X)=  a\cdot \log(L_R(Nuc)) +b
\end{equation}
where the coefficients ``a'' and ``b'' have been determined separately
for the whole sample (All) and for the Sy1 and NLSy1 galaxies alone,
respectively (Tab. \ref{ab}). The consistency of the relations is
confirmed by the Spearman Rank correlation coefficient, $\rho$, which is
higher than 0.7.  From the plot and the table it is obvious that Sy1
and NLSy1 behave in the same way.  The scatter is about $\pm$0.5
dex. There are (at least) two obvious reasons for this scatter:
first, optical and X-ray fluxes have not been measured simultaneously,
and due to the known variability of Seyferts a factor of a few can
easily be accounted for. Second, the X-ray fluxes have been
computed on the basis of the total measured counts and an {\em assumed} power
law slope.  It would have been more correct (though not possible in
most cases due to photon statistics) to fit also the power law slope,
and thus obtain a more accurate X-ray flux.
One may think that the $H_{\alpha}$ line may produce a
non-negligible contribution to the continuum emission in the
R-band.  For testing this hypothesis, the template spectrum of a Sy1
has been convolved with the transmission curve of the R-band
filter. We computed the dilution factor from H$_{\alpha}$, for
different [NII]/H$_{\alpha}$ ratios and for different
EW(H$_{\alpha}$), also taking into account that H$_{\alpha}$ may be
enhanced by galaxy interaction. The dilution factor is not more than
1\% and thus being negligible.\\

{\it Radio vs. X-ray emission.}
We also find that the X-ray emission correlates well with the radio
continuum emission. As in the previous case, the
relation is linear, and has been fit in the form
\begin{equation}
\log(L_{FIRST}(Nuc))= c \cdot \log(L_X) +d
\end{equation}
The coefficients ``c'' and ``d'' are listed in columns 4 and 5 of
Tab. \ref{ab}. Here we used only those 21 galaxies for which we have
information about nuclear radio emission. As previously, the linear
regression has been computed for all objects, and for the Sy1 (14
galaxies) and NLSy1 (7 galaxies) independently. Due to the small size
of the two subsamples the errors are quite high and the Spearman Rank
correlation coefficient in not really reliable.  Nevertheless, it
appears that NLSy1 may have a steeper slope than the Sy1 galaxies.\\
\begin{figure}[th]
\begin{center}
\hspace{-0.5cm}
 \vbox{\psfig{figure=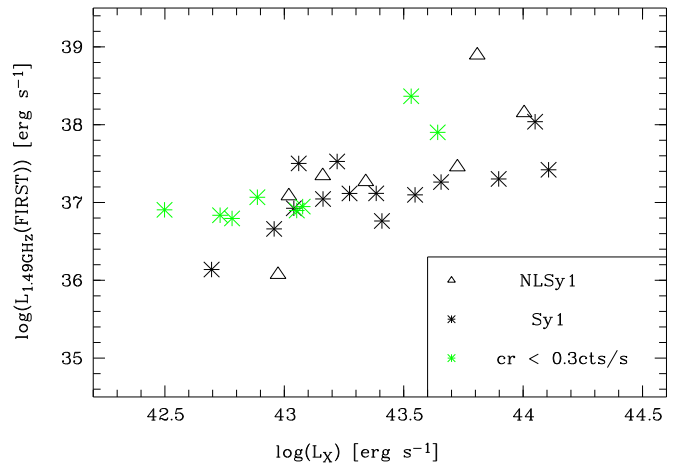,width=8.8cm,angle=-90,clip=}}\par
 \caption{Radio versus X-ray luminosity for the sample.  Symbols are
 the same as in Fig. \ref{lpsflx}. AGN with lower count rate
 than those we considered (grey asterisks) are also plotted, with
 the intention to show that our relation is not driven by the
 Malmquist bias.}
\label{lxlr}
\end{center}
\end{figure}

\begin{figure}[th]
\begin{center}
 \vbox{\psfig{figure=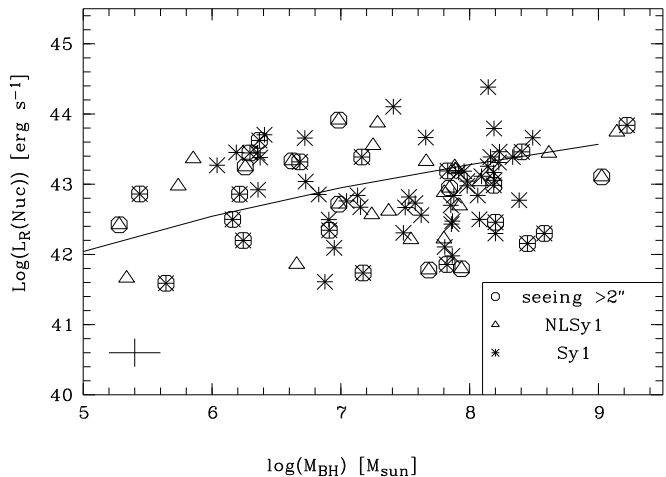,width=8.8cm,angle=-90,clip=}}\par
 \caption{R-band luminosity versus BH mass for the sample.
   Symbols are the same as in Fig. \ref{lpsflx}. At bottom
   left typical errors are shown. The plotted line is
   not a fit, but the predicted relationship for an optically thick
   accretion disk with $\dot{M}$ = 0.1 $\dot{M}_{Edd}$.}
\label{lpsflogmbh}
\end{center}
\end{figure}

{\it Optical vs BH masses.}
The optical luminosity of the accretion disk is also correlating with
the BH mass computed according to \S 3.2 (Fig. \ref{lpsflogmbh}).  The
solid line indicates the predicted relationship for an optically thick
accretion disk around a non-rotating BH (Shakura \& Sunyaev 1973),
where for a given BH mass the R band flux depends on the temperature
(and thus accretion rate $\dot{M}$) of the object {\bf ($\dot{M}=0.1
\dot{M}_{Edd}$} chosen here). The curvature of this line is basically
a measure of the differing bolometric correction for varying BH mass
(and thus temperature).


\section {Conclusions}

We find that the optical luminosity of the nuclei
of Seyfert galaxies correlates with the power-law X-ray emission,
the radio emission from the core as well as the Black Hole mass.

These relations are not driven by the Malmquist bias, which is
introduced whenever a sample is selected via a flux-limit.  In fact, the
additional 9 Sy1 galaxies with cr $<$ 0.3 cts/s and NVSS/FIRST
counterpart confirm the general trend (see Fig. \ref{lxlr}, grey
asterisks.)
   
The correlation of the optical nuclear luminosity with the X-ray
luminosity could potentially be understood if the canonical
blackbody-like continuum emission from the accretion disk is enhanced
due to irradiation by soft X-rays, so that an increase in the X-ray
flux would create a larger optical flux. This is a well-known effect
in X-ray binaries, though it has not been possible so far to quantify
this relation, as it depends on too many parameters (e.g. X-ray
spectrum of the irradiating source, temperature and ionization state
of the disk, geometry of the disk).

 Kriss \etal (1982) and Koratkar \etal (1995) found a strong correlation
between the soft X-ray luminosity and the broad $H_\beta$ and $H_\alpha$ 
line luminosity, respectively, in quasars and Sy1 galaxies.
It has been argued that an increased accretion disc flux would
not only cause a larger photoionization (and thus H$_{\alpha}$ flux),
but also provide more seed photons which would be up-scattered to
produce the power law X-ray spectrum, and thus the X-ray flux.
This effect is unlikely to explain our correlation, since
 simulations show that the contribution of the H$_{\alpha}$
emission line to the $R$ band flux is less than 1\%.

Including the correlation with the radio luminosity leads to the
consideration of a less direct, yet at least as plausible, coupling
between the luminosities in the different wavelengths, namely via the
electron corona which is supposed to surround the accretion disk. The
electron temperature distribution is expected to depend on the
accretion disk properties, primarily the accretion rate and disk
temperature. On the other hand, the temperature and density of the
corona determine the efficiency of the Compton scattering, and thus
directly determine the slope and power of the X-ray spectrum.  Thus, a
change in e.g. the accretion rate would not only lead to a varying
optical emission, but via the corona also to a varying X-ray and radio
emission.  The physical relation between the processes which produce
the emission in the X-ray and radio remains at the level of
speculation. In microquasars, the stellar-binary analogs of jet
emitting objects in our Galaxy, Rau \& Greiner (2002) have recently
found a similar correlation between the strength of the radio emission
and the X-ray power law slope. Owing to the smaller timescales involved
in microquasars (due to their smaller black hole mass), a whole
continuum of jet strengths has been observed from bright, spatially
resolved ejections over ``baby-jets'' down to quasi-steady radio
(synchrotron) emission.  Adopting this similarity, one could argue
that the core radio emission in Seyfert galaxies witnesses matter
outflow from the central black hole.  Objects with higher activity at
X-ray wavelengths, i.e. higher accretion rate, may thus also show
higher mass outflows from their central engine, thus implying a larger
radio flux.  Indeed, this is what has been found when comparing the
X-ray and radio fluxes of galactic black hole binaries in the low/hard
state: the radio flux scales as X-ray flux to the power of 0.7 (Gallo
\etal\ 2003).  Surprisingly, this is (within the errors) the same
correlation as we find (coefficient $c$ in Tab. \ref{ab}).  Certainly,
further investigations are necessary to understand the processes
involved and their interrelation, but it is encouraging that seemingly
the same processes are at work in AGN and galactic black hole
binaries.

\section{acknowledgements}

MS is grateful to G\"unther Hasinger and the AIP (Potsdam) for the
scientific and human support during the Ph.D. time.  MS also thanks
A. Merloni, D. Pierini and M. Radovich for their help and
discussion. The authors thank the anonymous referee for the useful
comments that improved the final version of this paper.  We made use
of the publicly available NVSS and FIRST radio catalogues.


\begin{thebibliography}{}

\bibitem[]{bks97}Bahcall, J.N. \etal, 1997, ApJ, 479, 642

\bibitem[]{bwg02}Becker, R.H., White, R.L., Helfand, D.J., 1995, ApJ, 450, 559

\bibitem[]{ccg98}Condon J.J. \etal, 1998, AJ, 115, 1693

\bibitem[]{dl90}Dickey, J.M., Lockman, F.~J., 1990, ARAA, 28, 215

\bibitem[]{Fa01} Falcke, H. \etal, 2001, in ``Black Holes in Binaries and Galactic Nuclei'',ESO Workshop in Garching.
 
\bibitem[]{fm00} Ferrarese, L. \& Merrit, D., 2000, ApJ, 539, L9

\bibitem[]{gfp03}Gallo E., Fender, R.P., Pooley, G.G., 2003, MNRAS, 344, 60

\bibitem[]{gbb00}Gebhardt, K. \etal, 2000, ApJ, 539, L13 

\bibitem[]{ksn00}Kaspi, S. \etal, 2000, ApJ, 533, 631

\bibitem[]{kdh95}Koratkar, A. \etal, 1995, ApJ, 440, 132

\bibitem[]{kcr82}Kriss, G.A., Canizares, C.~R., Ricker, G.R., 1980, ApJ, 242, 492

\bibitem[]{kwj01}Kuhlbrodt, B., Wisotzki, L., Jahnke, K., 2001, in
``QSO hosts and their environments'', Granada, eds. J. Marquez \etal,
 Kluwer Acad. Press

\bibitem[]{mtr98}Magorrian, J. \etal, 1998, AJ, 115, 2285

\bibitem[]{mld01}McLure, R.J., Dunlop, J.S., 2001, MNRAS, 327, 199

\bibitem[]{mld02}McLure, R.J., Dunlop, J.S., 2002, MNRAS, 331, 795

\bibitem[]{ps95}Predehl, P., Schmitt, J.H.M.M., 1995, A\&A, 293, 889

\bibitem[]{rg03}Rau, A., Greiner, J., 2003, A\&A, 397, 711

\bibitem[]{ms02}Salvato, M., 2002, Ph.D. Thesis, University of Potsdam
 (Germany)

\bibitem[]{shl00}Schwope, A. \etal, 2000, Astron. Nach., 321, 1

\bibitem[]{ss73}Shakura, N.I. \& Sunyaev, R.A., 1973, A\&A, 24, 337

\bibitem[]{vvg01}V\' eron-Cetty, M.P., V\' eron, P., Gon\c calves, A.C., 2001, A\&A, 372, 730

\bibitem[]{vab99}Voges, W. \etal, 1999, A\&A, 349, 383

\bibitem[]{vi00}Virani, S.N., De Robertis, M.M, VanDalfsen, M.L., 2000, AJ, 120, 1739
\end{thebibliography}
\end{document}